\magnification=\magstep1
\hsize=120truemm
\vsize=186truemm
\baselineskip=12truept

\font\fontA=cmbx10 scaled \magstep1

\ 
\input epsf

\noindent\leftline{{\fontA Equilibrium and off-equilibrium simulations of
}} 

\leftline{{\fontA chiral-glass order in 
three-dimensional}}

\leftline{{\fontA 
 Heisenberg spin glasses
}}

\par\bigskip\medskip
\noindent
\leftline{H. Kawamura$^a$ and K. Hukushima$^b$}
\bigskip
\leftline{$^a$Faculty of Engineering and Design, 
Kyoto Institute of Technology,   
}
\leftline{Kyoto 606-8585, ~JAPAN}
\leftline{$^b$ISSP, University of Tokyo, Tokyo,
106-8666, ~JAPAN}
\bigskip\bigskip\medskip\noindent
{\bf Abstract.}  
Spin-glass and chiral-glass orderings in three-dimensional 
Heisenberg
spin glasses are
studied both by equilibrium and 
off-equilibrium Monte Carlo simulations.
Fully isotropic model is found to exhibit 
a finite-temperature
chiral-glass transition without the conventional spin-glass order.
Although chirality is an Ising-like quantity from symmetry, 
universality class of the chiral-glass transition appears to
be different from
that of the standard Ising spin glass. 
In the off-equilibrium simulation, while the spin 
autocorrelation exhibits only
an interrupted aging, the chirality autocorrelation
persists to exhibit a pronounced aging
effect reminiscent of the one observed
in the mean-field model.
Effects of random magnetic anisotropy is also studied
by the off-equilibrium simulation, in which
asymptotic mixing of the spin and the chirality
is observed.
\par\bigskip\medskip

\noindent
\S 1. {\bf Introduction\/}
\par\medskip

Ordering of complex systems has 
attracted  interest of 
researchers working in the field of
numerical simulations.
Well-known examples of such complex systems 
may be a variety of `glassy' systems including window glasses,
orientational glasses of molecular crystals, 
vortex glasses in superconductors and
spin-glass magnets.
Often, in the
dynamics of such complex systems,  characteristic slow relaxation 
is known to occur.
It has been a great challenge for  researchers  
to clarify the nature and the origin of these slow dynamics, as
well as to
get fully equilibrium properties by overcoming
the slow relaxation.
In particular, {\it spin glasses\/} 
are the most extensively studied typical
model system, 
for which numerous analytical, numerical and experimental
works have been made [1].

Studies on spin glasses now have more than twenty years of history.
Main focus of earlier studies was put on obtaining the
equilibrium properties of the spin-glass ordering.
Owing to extensive experimental studies, it now seems
well-established that the spin-glass magnets exhibit 
an equilibrium  phase transition 
at a finite temperature [1]. From theoretical side, 
there now seems to be a consensus 
that the lower critical dimension (LCD) 
of an Ising 
spin glass 
is between $d=2$ and 3, while the LCD of
vector spin glasses 
is greater than $d=3$.
In other words, at $d=3$, only an anisotropic
Ising spin glass exhibits
an equilibrium spin-glass transition at a finite temperature
[2,3], whereas 
isotropic 
Heisenberg 
spin glass exhibits only a zero-temperature transition [4-7].
Meanwhile, it 
has been known that the magnetic interactions in many of 
real spin-glass materials are 
Heisenberg-like, in the sense that the magnetic anisotropy is
much weaker than the exchange energy.
Apparently, there is a puzzle here: How can one reconcile the
absence of the spin-glass order in an isotropic 
Heisenberg spin glass with the experimental observation? 
While weak magnetic anisotropy inherent to real materials
is often invoked to explain this apparent discrepancy,
it remains puzzling that no
detectable sign of
Heisenberg-to-Ising crossover has been observed in experiments
which is usually expected to occur if the observed
spin-glass transition 
is caused by the weak magnetic anisotropy [1,4].

In order to solve this apparent puzzle, a  chirality mechanism of
experimentally observed spin-glass transitions
was recently proposed by one of the authors [6],  on the assumption 
that an isotropic 3D Heisenberg
spin glass 
exhibited a finite-temperature 
{\it chiral-glass\/} transition without the conventional spin-glass 
order, in which only spin-reflection 
symmetry was broken with
preserving spin-rotation symmetry. `Chirality' is an Ising-like
multispin variable representing the
sense or the handedness of the noncoplanar spin structures 
(more detailed definition will
be given below).
It was argued that, in real spin-glass magnets, 
the spin and the chirality were ``mixed'' due to the weak
magnetic anisotropy and the chiral-glass transition
was then ``revealed'' via anomaly in
experimentally accessible quantities. 
Meanwhile, 
theoretical question whether there really occurs such 
finite-temperature chiral-glass transition in an isotropic 3D
Heisenberg spin glass,
a crucial assumption of the chirality mechanism, 
remains somewhat inconclusive [6,7].

More recently, 
there arose a growing interest both theoretically and
experimentally in the {\it off-equilibrium\/} dynamical properties of 
spin glasses. In particular,
aging phenomena observed
in many spin glasses [8]
have attracted  attention of researchers [9,10].
Unlike systems in thermal equilibrium,  relaxation of physical
quantities depends not only on the observation time $t$ but also on 
the waiting time $t_w$, {\it i.e.\/}, how long
one waits at a given state before the measurements. 
Recent studies  have revealed that
the off-equilibrium dynamics  in the spin-glass  state generally 
has two
characteristic time regimes [9-11]. One is a short-time 
regime, $t_0<<t<<t_w$ ($t_0$ is a microscopic time scale), 
called `quasi-equilibrium regime', and the
other is a long-time regime, $t>>t_w$, called `aging regime' or
`out-of-equilibrium regime'. In the quasi-equilibrium regime,
the relaxation is stationary and the fluctuation-dissipation
theorem (FDT) holds. The 
autocorrelation function at  times $t_w$ and $t+t_w$
is expected to behave as
$$C(t_w,t+t_w)\approx q^{{\rm EA}}+{C\over t^\lambda }\rightarrow
q^{{\rm EA}},\eqno(1)$$
where $q^{{\rm EA}}$ is the equilibrium Edwards-Anderson order
parameter.
In the aging regime, the relaxation becomes non-stationary, 
FDT broken,
and the autocorrelation function decays to zero
as $t\rightarrow \infty $ for fixed 
$t_w$.

On theoretical side, both analytical and numerical studies 
of off-equilibrium dynamics of spin glasses have so far been
limited  to   {\it Ising-like\/} models, including the
Edwards-Anderson (EA) 
model with short-range interaction [12-14]
or the mean-field models with long-range interaction [11,15-17].  
Although these analyses on Ising-like  models
succeeded in reproducing some of the features of
experimental results, 
it is clearly desirable to study the dynamical
properties of {\it Heisenberg-like\/} spin-glass models
to make a direct link between theory and experiment.

In the present article, we report on our recent results of 
equilibrium as well as off-equilibrium Monte Carlo simulations on  
isotropic and  anisotropic  3D Heisenberg
spin glasses.
Ordering properties of both
the spin and the chirality will be studied, aimed at 
testing the
validity of the proposed 
chirality scenario of spin-glass transitions.
We note that Monte Carlo simulation is particularly suited to this
purpose,
since, at the moment, chirality itself is not directly 
measurable experimentally. By contrast, in numerical simulations, 
it is quite
straightforward to measure the chirality.
\par\bigskip
\noindent
\S 2. {\bf Chirality}
\medskip
Frustration in vector spin systems such as the {\it XY\/} and
Heisenberg models
often causes {\it noncollinear\/} or {\it noncoplanar\/} spin
structures. Such noncollinear or noncoplanar orderings give rise to
a nontrivial chirality.
In the case of 
two-component {\it XY\/} spins, 
one may define a local chirality for the two neighboring spins
at the $i$- and $j$-th sites by
$\kappa _{ij}={\bf S}_i\times {\bf S}_j\mid _z=
S_i^xS_j^y-S_i^yS_j^x$.
When the spin configuration in the ordered state is noncollinear,
the local chirality  $\kappa _{ij}$ defined above takes a nonzero value.

In the case of three-component Heisenberg spins, 
three spins are necessary to define a scalar chirality.
Thus, in the Heisenberg case, 
one may define a local scalar chirality  
for three neighboring spins (spin triad) at the 
$i$-, $j$- and $k$-th sites by
$\chi _{ijk}={\bf S}_i\cdot {\bf S}_j\times {\bf S}_k$.
It  takes a nonzero value for any noncoplanar spin 
configurations but
vanishes for any planar spin configurations.
Note that, in either case,  the chirality is a pseudoscalar in the
sense that it
is invariant under global spin rotation but changes sign under 
global spin reflection.
Possible chiral ordering is related with a
breaking of the reflection symmetry with preserving the rotation
symmetry.

The model we simulate is the classical Heisenberg
model on 
a simple cubic lattice 
with the nearest-neighbor  random Gaussian couplings, $J_{ij}$ 
and $D_{ij}^{\mu \nu }$,
defined by the Hamiltonian
$${\cal H}=-\sum_{<ij>} (J_{ij}{\mit\bf S}_i\cdot
{\mit\bf S}_j+D_{ij}^{\mu \nu }S_i^\mu S_i^\nu ),\eqno(2)$$
where ${\bf S}_i$ 
=$(S_i^x,S_i^y,S_i^z)$ 
is a three-component 
unit vector, and 
the sum runs over all nearest-neighbor pairs 
with $N=L\times L\times L$ spins.   
$J_{ij}$ is the isotropic exchange coupling 
with zero mean and variance $J$,  while $D_{ij}^{\mu \nu}\ 
(\mu, \nu =x,y,z)$  is the random magnetic anisotropy 
with zero mean and variance $D$ which is assumed to be
symmetric and traceless,
$D_{ij}^{\mu \nu }=D_{ij}^{\nu \mu }$ and $\sum _\mu 
D_{ij}^{\mu \mu }=0$.

We define the local chirality at the $i$-th site and in the $\mu $-th 
direction, 
$\chi _{i\mu }$, for three Heisenberg spins by,
$$\chi _{i\mu }={\bf S}_{i+\hat {\bf e}_\mu }\cdot ({\bf S}_i\times 
{\bf S}
_{i-\hat {\bf e}_\mu }),\eqno(3)$$
where $\hat {\bf e}_\mu \ (\mu =x,y,z)$ 
denotes a unit lattice vector along the
$\mu\/$-axis.

\par\bigskip\medskip
\noindent
\S 3. {\bf Equilibrium simulations}
\medskip
First, we report on our {\it equilibrium\/} Monte Carlo simulation
of a fully isotropic 3D Heisenberg spin glass defined by eq.(2) 
with $D=0$.
Monte Carlo simulation is performed based on
an `extended ensemble' method
recently developed by Hukushima and Nemoto [18], where 
the whole
configurations at two neighboring temperatures of the same sample
are occasionally exchanged  with the system remaining 
at equilibrium. 
By this method, we 
succeeded in equilibrating the system down to the temperature
considerably lower than those attained in the previous simulations.
We run  in parallel two independent replicas 
with the same bond realization
and compute an overlap between the chiral variables in the two
replicas,
$$q_{\chi } =  {1\over 3N}\sum _{i,\mu} \chi _{i\mu} ^{\{1\}}
\chi _{i\mu} ^{\{2\}}.\eqno(4)$$ 
In terms of this chiral overlap, $q_{\chi} $, the chiral-glass
order parameter, $q_{{\rm CG}}^{(2)}$, and
the Binder cumulant of the chirality, $g_{{\rm CG}}$,  
are calculated by
$$q_{{\rm CG}}^{(2)}=[<q_\chi ^2>]^2,\eqno(5)$$
$$g_{{\rm CG}}=  {1\over 2}
(3- {[<q_\chi ^4>]\over [<q_\chi ^2>]^2}),\eqno(6)$$ 
where $<\cdots >$ represents the thermal average and $[\cdots ]$ 
represents the average over bond disorder.
At the possible chiral-glass
transition point,
curves of $g_{{\rm CG}}$ against $T$ 
for different $L$ should merge or cross 
asymptotically for large $L$.

For the Heisenberg spin,
one can introduce an appropriate Binder cumulant in terms of 
a tensor overlap $q_{\mu \nu}\ (\mu ,\nu =x,y,z)$
which has $3^2=9$ independent components, 
$$q_{\mu \nu}\equiv {1\over N}
\sum _iS_{i\mu}^{\{1\}}S_{i\nu}^{\{2\}},
\ \ (\mu,\nu =x,y,z),\eqno(7)$$
via the relation,
$$g_{{\rm SG}}={1\over 2}
(11-9 {\sum _{\mu,\nu,\delta,\rho} [<q_{\mu \nu}^2
q_{\delta \rho}^2>]\over (\sum _{\mu,\nu} 
[<q_{\mu \nu}^2>])^2}).\eqno(8)$$ 
%

The lattice sizes studied are
$L=6,8,10,12,16$ with periodic boundary conditions.
In the case of $L=12$, for example, we prepare 50
temperature points distributed in the range [0.08J, 0.25J] 
for a given sample, 
and perform $4.7\times 10^5$
exchanges  per temperature of the whole lattices
combined with the same number of standard 
single-spin-flip heat-bath sweeps.
For $L=12$, we equilibrate the system down to
the temperature $T/J=0.08J$, 
which is lower than the minimum
temperature attained previously.
Sample average is taken over 1500 ($L=6$), 1200 ($L=8$), 640 ($L=10$),
296 ($L=12$) and 32 ($L=16$) independent bond realizations.
Equilibration is checked by monitoring the 
stability of the results
against at least three-times longer runs for a subset of samples.
\par
The size and temperature dependence 
of the Binder cumulants of the spin and of the chirality,
$g_{{\rm SG}}$ and $g_{{\rm CG}}$, are shown in
Fig.1(a) and (b), respectively. As can be seen from Fig.1(a),
$g_{{\rm SG}}$ constantly decreases with increasing $L$ 
at all temperatures studied,
suggesting that 
the conventional spin-glass order occurs only at
zero temperature, consistent with the previous results [1,4-7].
A closer inspection of Fig.1(a), however, reveals that $g_{{\rm SG}}$ 
for larger lattices ($L=10,12,16$) exhibits an anomalous ``upturn'' 
around $T/J\sim 0.1-0.15$, suggesting that a change in the ordering
behavior occurs in this temperature range. 

\smallskip\centerline{
\epsfysize=4.5cm
\epsfxsize=5cm
\epsfbox{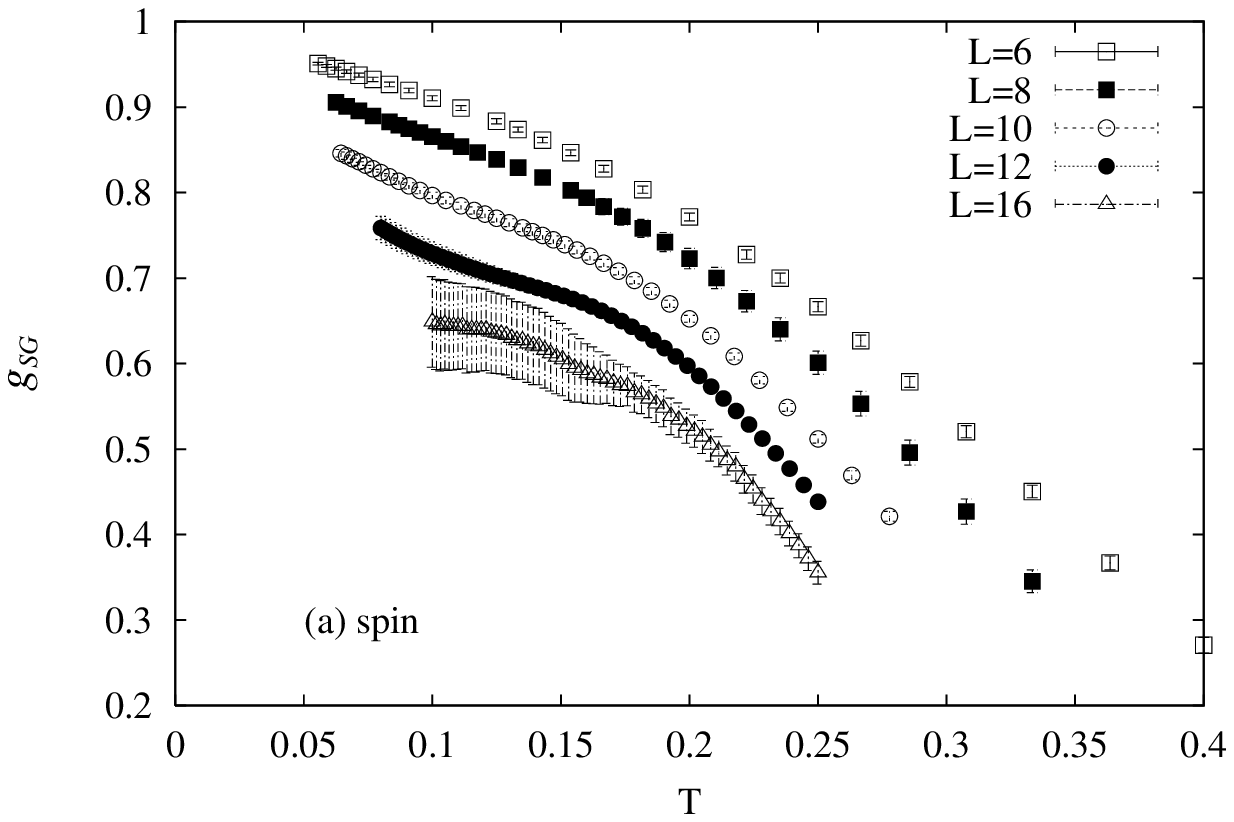}
\epsfysize=4.5cm
\epsfxsize=5cm
\epsfbox{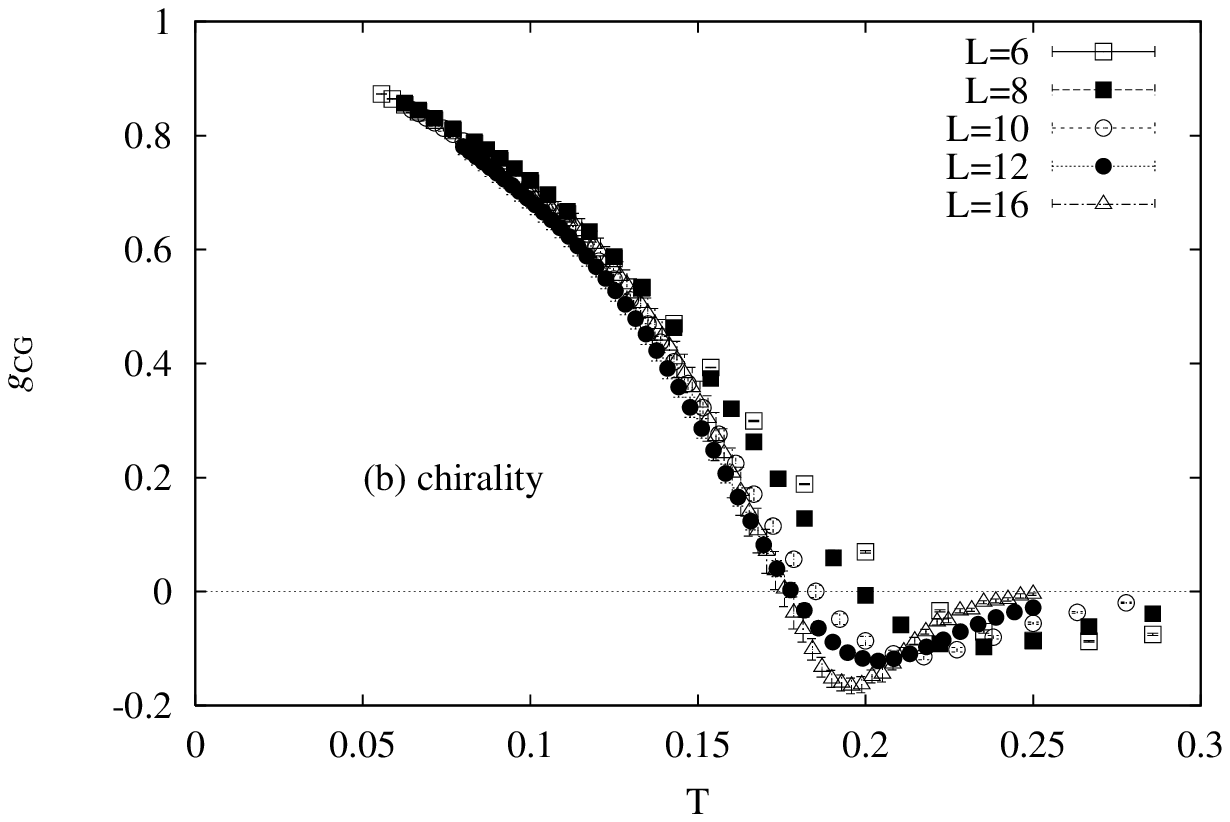}}
\item{FIG.1~} Temperature and size dependence of the Binder cumulants
of the spin (a) and  of the chirality (b) of a 3D  isotropic
Heisenberg spin glass.

As can be seen from Fig.1(b), 
the curves  of $g_{{\rm CG}}$ for different $L$ do not cross, but
show a tendency to merge for larger $L$
in the temperature range where the curves of $g_{{\rm SG}}$ exhibit
an anomalous upturn.
However, it is not possible to determine only 
from the data of $g_{{\rm CG}}$ whether
the system exhibits a finite-temperature  transition
into the chiral ordered state with some ``critical'' character, or
it exhibits only a zero-temperature transition with rapidly
growing correlation length. 

Thus, we also have tried the standard finite-size scaling analysis
for the chiral-glass order parameter $q_{{\rm CG}}^{(2)}$, 
by adjusting
the transition temperature $T_{{\rm CG}}$ and the exponents $\beta 
_{{\rm CG}}$ and $\nu _{{\rm CG}}$ as fitting parameters.
The best fit is obtained for 
$T_{{\rm CG}}/J\sim 0.176$, $\nu _{{\rm CG}}\sim 1.15$ and 
$\beta _{{\rm CG}}/\nu _{{\rm CG}}\sim 1.30$, 
with the associated $\chi^2_{\rm dof}$-value, $\chi^2_{\rm dof}\sim 5.3$; 
see Fig.2(a). 
If, on the other hand, the same data are fitted to the
finite-size scaling form expected for a
$T=0$ transition with nondegenerate ground state, 
{\it i.e.\/}, $T_{{\rm CG}}=0$ and $\beta _{{\rm CG}}/\nu _{{\rm CG}}
=0$, the best fit is obtained for 
$\nu _{{\rm CG}}\sim 3.2$, but with the associated 
$\chi ^2$-value, $\chi ^2\sim 29.5 $, which is 
significantly 
larger than the best $\chi ^2$-value obtained with assuming
$T_{{\rm CG}}>0$.
Thus, while our individual estimates of
$T_{{\rm CG}}$, $\nu _{{\rm CG}}$ and 
$\beta _{{\rm CG}}/\nu _{{\rm CG}}$
might not be
so accurate since we are trying here
to determine the three intercorrelated
fitting parameters 
simultaneously, our analysis strongly favors a finite-temperature
chiral-glass transition over a zero-temperature transition.
Indeed, this conclusion is corroborated with the results of the
dynamical simulation to be presented in the next section.

\smallskip\centerline{
\epsfysize=5cm
\epsfxsize=6cm
\epsfbox{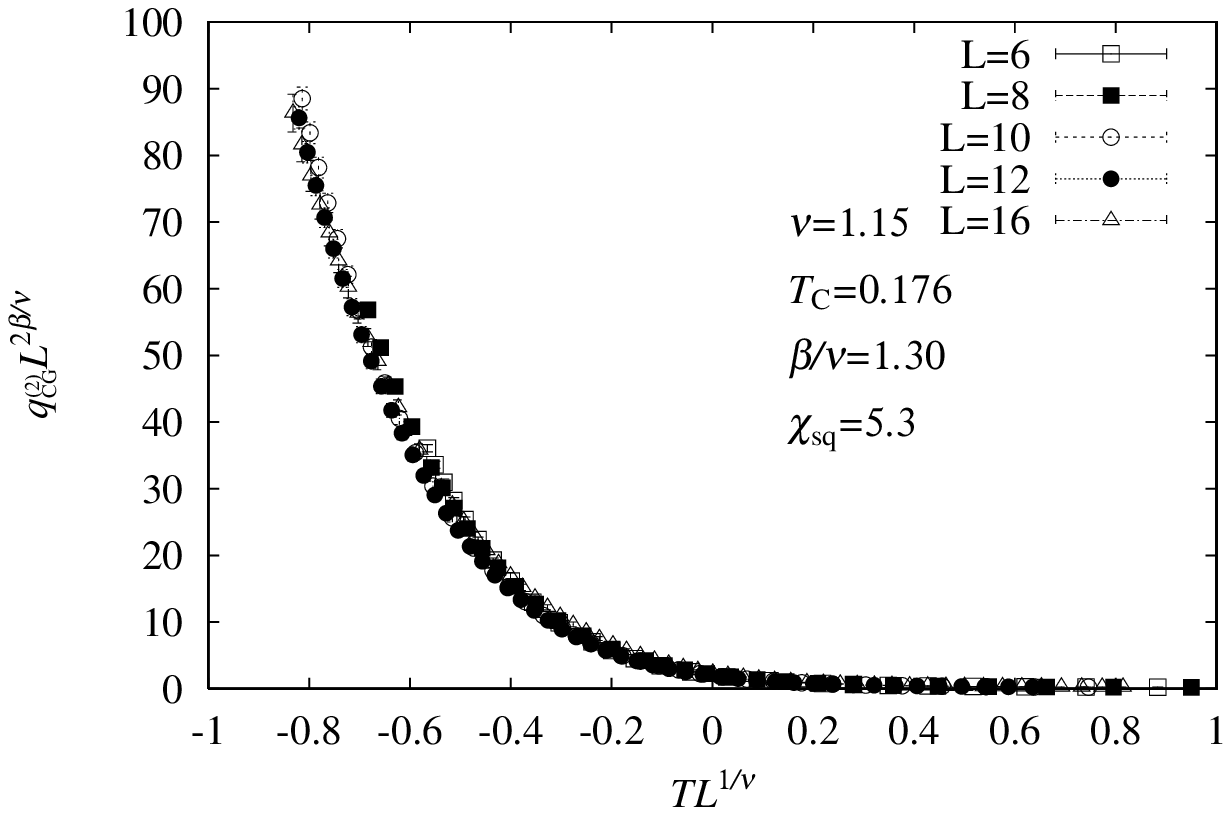}
}
\item{FIG.2~} Finite-size scaling plots of the chiral-glass
order parameter of a 3D isotropic Heisenberg
spin glass, with assuming a finite-temperature transition 
$T_{{\rm CG}}>0$.
The chiral-glass order parameter is divided by the magnitude of the
local chirality.[6,7]\smallskip

In Fig.3, we display the distribution function of the chiral-overlap 
defined by
$$P_{{\rm CG}}(\bar q_\chi )=[<\delta (q_\chi -\bar q_\chi )>],
\eqno(9)$$
calculated at a temperature $T/J=0.1$, below the estimated
chiral-glass transition temperature.
The shape of the calculated $P_{{\rm CG}}(q_\chi )$ is
somewhat different from the  one 
observed in the standard Ising-like models such as the
3D EA model or the mean-field
SK mode.
$P_{{\rm CG}}(q_\chi )$ has standard `side-peaks'
corresponding to the Edwards-Anderson order parameter 
$\pm q_{{\rm CG}}^{EA}$, which 
grow and sharpen with increasing $L$
as is usually the case in the spin-glass state.
In addition to the side peaks, 
a `central peak' at $q_\chi =0$ shows up for larger $L$, 
which also grows and sharpens with increasing $L$. This latter
aspect, {\it i.e.\/}, the existence of a central peak growing
and sharpening with the system size, is a peculiar feature of
the chiral-glass ordered phase never
observed in 
the EA model or the  SK model: It might  suggest a 
nontrivial structure in the phase space associated with the
chirality. 
This peculiar feature
is  reminiscent of the behavior characteristic of  some mean-field
models showing the so-called {\it 
one-step replica-symmetry breaking\/} (RSB).
Thus, it is  tempting to deduce that the 
chiral-glass phase of a 3D Heisenberg spin glass
has a character of such one-step RSB.
If the ordering is really of such type,  
no crossing of the Binder ratio needs to occur at $T=T_{{\rm CG}}$,
and our data of $g_{{\rm CG}}$ are consistent with
the scenario. Further studies are in progress to clarify   
the situation.

\smallskip\centerline{
\epsfxsize=8cm
\epsfbox{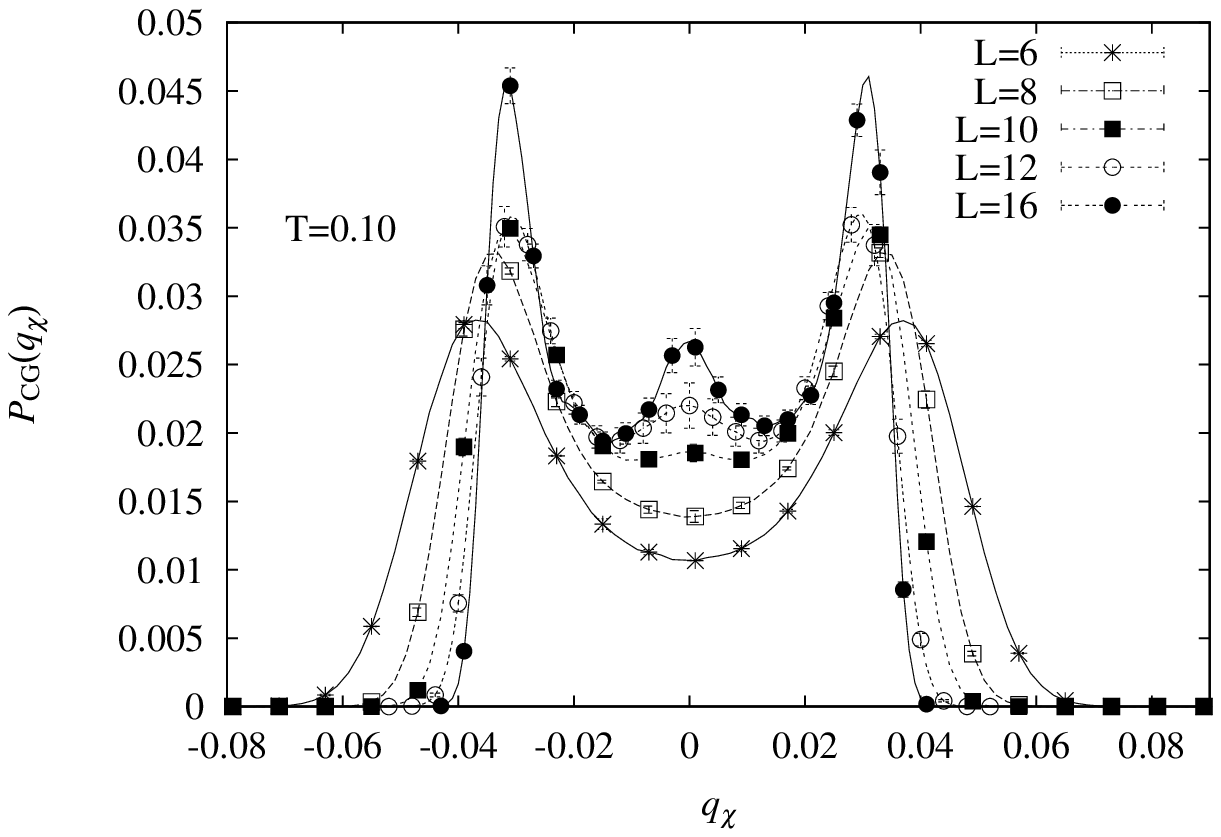}
}
\item{FIG.3~} Chiral-overlap distribution function of a 3D isotropic
Heisenberg spin glass
below $T_{{\rm CG}}$. The temperature is 
$T/J=0.1$.

\par\bigskip\medskip
\noindent
\S 4. {\bf Off-equilibrium simulations}
\medskip

In this section, we report on the results of {\it off-equilibrium\/}
Monte Carlo simulations both on isotropic ($D=0$) and anisotropic
$(D>0)$ models. Unlike the case of
equilibrium simulation, 
the system here is never in full thermal equilibrium. Recent
studies have revealed that one can still get
many useful information from
such off-equilibrium simulations, even including certain
{\it equilibrium\/} properties. The quantities we are mainly 
interested here are
the spin and  chirality
autocorrelation functions 
defined by
$$C_s(t_w,t+t_w)={1\over N}
\sum _i[<\vec S_i(t_w)\cdot \vec S_i(t+t_w)>],\eqno(10)$$
$$C_\chi (t_w,t+t_w)={1\over 3N}\sum _{i,\mu }[<\chi _{i\mu }(t_w) 
\chi _{i\mu }(t+t_w)>].\eqno(11)$$

Monte Carlo simulation is performed based on the standard single
spin-flip heat-bath method.
Starting from completely random initial
configurations, the system is  
quenched to a working temperature.
Total of about $3\times 10^5$ 
Monte Carlo steps per spin [MCS] are generated
in each run.
Sample average is taken over 30-120 independent bond realizations, 
four independent runs being made using different
spin initial conditions and different sequences of random numbers
for each sample.
The lattice size  mainly studied is $L=16$ with periodic
boundary conditions, while in some cases lattices with  $L=12$
and $24$ are also studied.
\par\smallskip

Let us begin with the fully isotropic case, $D=0$.
The spin and chirality autocorrelation functions 
at a low temperature $T/J=0.05$ are
shown in Fig.4 as a function of $t$.
For larger $t_w$,
the curves of the spin autocorrelation function  $C_s$ 
come on top of each other
in the long-time
regime, indicating that the stationary
relaxation is recovered and
aging  is
interrupted. This behavior has been expected because the 3D
Heisenberg spin glass has
no standard spin-glass order [1,4-7]. 
Similar interrupted aging was observed 
in the 2D Ising
spin glass which did not have an equilibrium spin-glass order [13]. 
By contrast, the chiral 
autocorrelation function $C_{\chi }$ 
shows an entirely different behavior:
Following the initial decay, it exhibits a  clear plateau
around $t\sim t_w$ and then drops sharply for $t>t_w$. 
It also shows an eminent aging effect, namely, as one waits longer,
the relaxation becomes slower and the plateau-like behavior at
$t\sim t_w$ becomes more pronounced. 

\smallskip\centerline{
\epsfysize=4.5cm
\epsfxsize=5cm
\epsfbox{fig4a.ps}
\epsfysize=4.5cm
\epsfxsize=5cm
\epsfbox{fig4b.ps}
}
\item{FIG.4~} Spin (a) and chirality (b) autocorrelation functions
of a 3D  isotropic Heisenberg spin glass at a temperature $T/J=0.05$
plotted versus 
$\log _{10}t$ for various waiting times $t_w$. The lattice size
is $L=16$ averaged over 66 samples.
\smallskip
\smallskip\centerline{
\epsfxsize=5cm
\epsfysize=4.5cm
\epsfbox{fig5a.ps}
\epsfxsize=5cm
\epsfysize=4.5cm
\epsfbox{fig5b.ps}
}
\item{FIG.5~} The same data as in Fig.4, 
but plotted versus $\log _{10}(t/t_w)$.
\smallskip

In Fig.5, $C_s$ and $C_\chi $  are replotted
as a function of the
scaled time $t/t_w$.
Reflecting its interrupted aging, the curves of $C_s$ 
for larger $t_w$ now lie below the ones
for smaller $t_w$
({\it subaging\/}). 
By contrast, the curves of $C_\chi $ for
various $t_w$ cross around $t/t_w\sim 1$, and at $t>t_w$, 
the data for larger $t_w$ lie {\it above\/} the ones 
for smaller $t_w$ ({\it superaging\/}).
Such superaging behavior of $C_\chi $ means 
that the aging in chirality is
more enhanced than the one expected from the naive 
$t/t_w$-scaling. Note that, although the
chirality is an Ising-like variable from  symmetry,
the observed superaging behavior is in  contrast to the aging
behavior of the 3D EA Ising model
which was found 
to satisfy a
good $t/t_w$-scaling in the aging regime [13]. 
It should also be noticed that 
the plateau-like behavior observed here has been 
hardly noticeable
in  simulations of the 3D EA Ising model.
Rather, the behavior of $C_\chi $ observed here
is reminiscent of the one observed in
the mean-field model
such as the SK 
model [15-17]. This correspondence might suggest
that an effective interaction between the chiralities is long-ranged.
Indeed, at least in case of {\it XY\/} spins, 
the interaction between
chiralities is known to be
Coulombic [19], while such analytical information is not available
for Heisenberg spins.

While the 
plateau-like behavior observed in $C_\chi $ is already suggestive 
of a nonzero {\it chiral\/} Edwards-Anderson order parameter, 
$q_{{\rm CG}}^{{\rm EA}}>0$,  more quantitative
analysis similar to the one recently done by Parisi {\it et al\/}
for the 4D Ising spin glass [14] is performed  
to extract $q_{{\rm CG}}^{{\rm EA}}$ from the data of
$C_\chi $ in the quasi-equilibrium regime.
Finiteness of $q_{{\rm CG}}^{{\rm EA}}$ is also visible in a
log-log plot of
$C_\chi $ versus $t$ as shown in the inset 
of Fig.6, where the data show a clear upward curvature.
We extract $q_{{\rm CG}}^{{\rm EA}}$ by fitting 
the data of $C_\chi $ for
$t_w=3\times 10^5$
to the power-law form of eq.(1) in the time range
$40\leq t\leq 3,000$
satisfying $t/t_w\leq 0.01$. 
Stability of the result has been 
checked by examining the robustness of
the result against the change in the value of $t_w$ and the 
time window used in the fit. 
The obtained $q_{{\rm CG}}^{{\rm EA}}$,
plotted as a function of temperature in Fig.6,  
clearly indicates the
occurrence of a finite-temperature chiral-glass transition
at $T_{{\rm CG}}/J=0.157\pm 0.01$ with the associated 
order-parameter exponent
$\beta _{{\rm CG}}=1.1\pm 0.1$. 
The size dependence turns out to be rather small,
although the mean values of $q_{{\rm CG}}^{{\rm EA}}$
tend to slightly
increase around $T_{{\rm CG}}$ with increasing $L$.
Since both  finite-size effect and finite-$t_w$ effect
tend to underestimate  $q_{{\rm CG}}^{{\rm EA}}$, one may
regard the present result as a rather strong
evidence of the occurrence of a finite-temperature 
chiral-glass transition. 

\smallskip\centerline{
\epsfxsize=6cm
\epsfbox{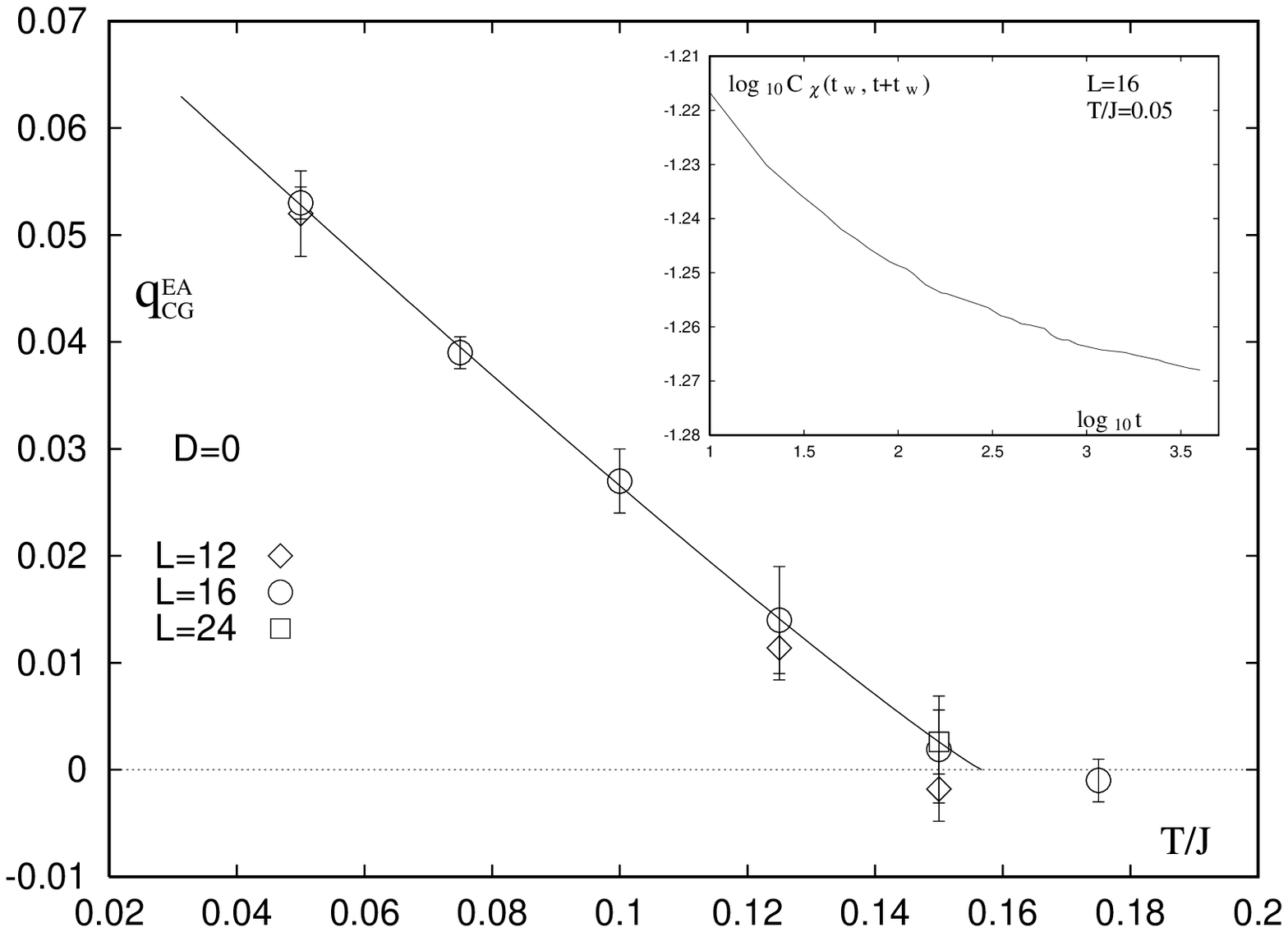}
}
\item{FIG.6~} Temperature dependence of the Edwards-Anderson order
parameter of the chirality of a 3D  isotropic Heisenberg spin glass.
The data are averaged over 30-120 samples.
Inset exhibits the log-log plot of the
$t$-dependence of the
chirality  autocorrelation function in the quasi-equilibrium regime
for $L=16$ and 
$t_w=3\times 10^5$. 
\smallskip

If one compares the present estimate with the static result
in the preceding section, the estimated transition temperature
is a bit lower than, but is roughly consistent with the 
static estimate
$T_{{\rm CG}}/J\sim 0.17$, while the exponent $\beta _{{\rm CG}}$
is somewhat smaller than the static estimate, $\beta _{{\rm CG}}\sim
1.5$. In any case,
the obtained
exponent $\beta _{{\rm CG}}\sim 1.1$ (or still larger value
from the static estimate)
is considerably
larger than the value of the 3D EA model $\beta \sim 0.5$ [1-3], 
and is rather close to the value of the mean-field model $\beta =1$.
This suggests that the universality class of the chiral-glass
transition of the 3D Heisenberg spin glass might be 
different from that of the standard 3D Ising spin glass.
According to the chirality mechanism,
the criticality of real
spin-glass transitions 
should be the same as that of the chiral-glass
transition of an isotropic Heisenberg spin glass, so long as the
magnitude of  random anisotropy is not too strong.
If one tentatively accepts this scenario, 
the present result opens up a new interesting possibility
that the universality class of many of real spin-glass transitions 
might differ from that of the standard Ising
spin glass, contrary to common belief.

In the presence of
weak  anisotropy $D>0$, chirality scenario predicts at the static level that
the transition behavior of  chirality remains essentially 
the same
as in the isotropic case, whereas the spin is 
mixed into the chirality,
asymptotically
showing the same transition behavior as the chirality [6].
In order to see whether such ``spin-chirality mixing'' occurs  in 
the off-equilibrium dynamics, 
further dynamical simulations are performed for the models with 
random anisotropies $D/J=0.01\sim 1$.
While
chirality exhibits essentially the same dynamical behavior
as in the isotropic case (not shown here),
the  behavior
of spin at $t>t_w$ 
changed significantly in the presence
of  anisotropy. 
As an example, the spin autocorrelation in the case of 
weak anisotropy $D/J=0.01$ is shown in Fig.7.
Even for such small anisotropy,
spin is found to show
{\it superaging\/} behavior asymptotically
at $t>>t_w$ similar to that
of the chirality in {\it zero\/} and weak anisotropies, 
demonstrating the spin-chirality mixing.

\smallskip\centerline{
\epsfxsize=6cm
\epsfbox{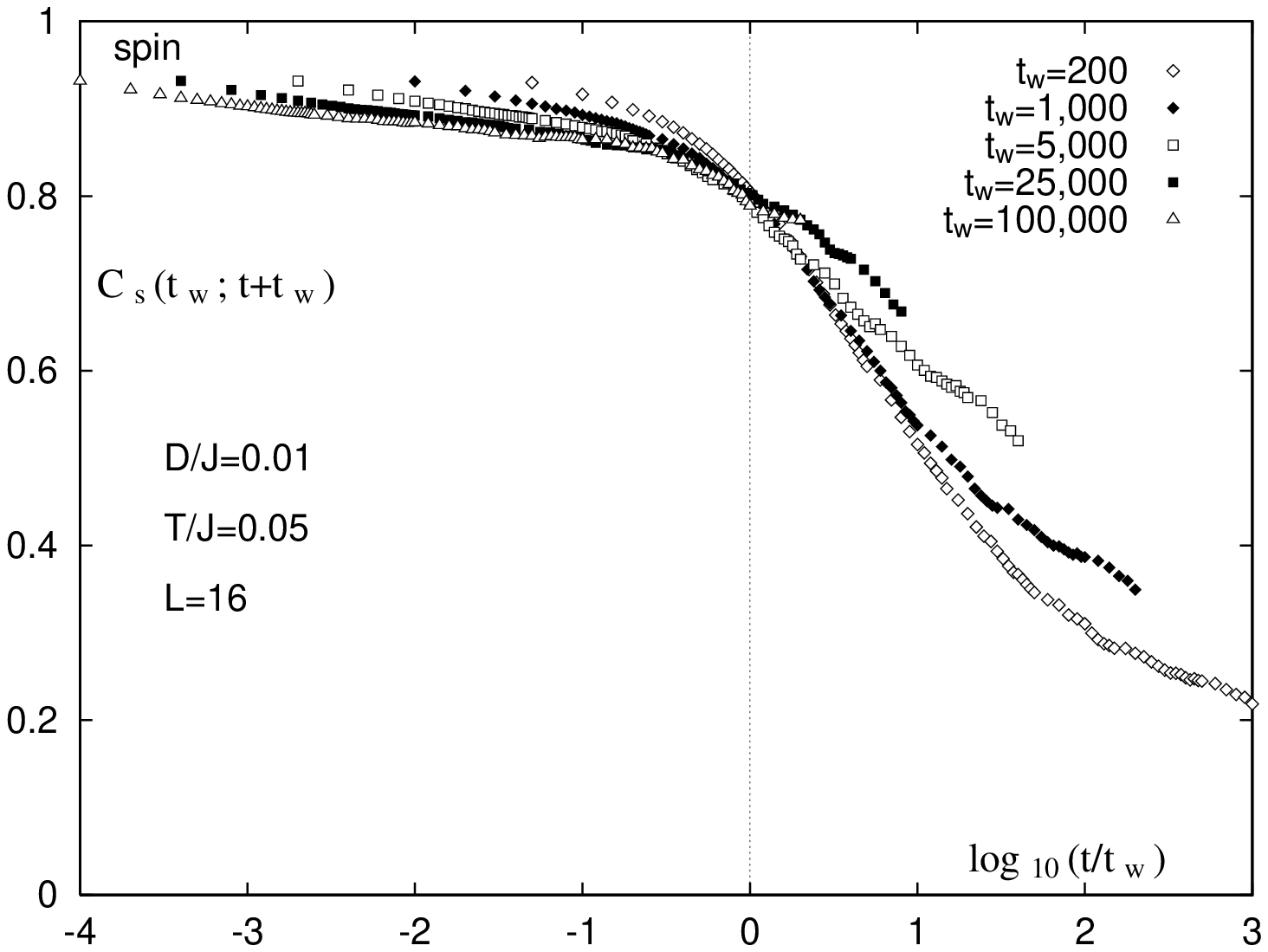}
}
\item{FIG.7~} Spin autocorrelation function
of the weakly anisotropic 3D Heisenberg spin glass 
with $D/J=0.01$ plotted versus  $\log_{10}(t/t_w)$. 
The lattice size is 
$L=16$ averaged over 60 samples and the temperature is $T/J=0.05$. 
\smallskip

\smallskip\centerline{
\epsfxsize=6cm
\epsfbox{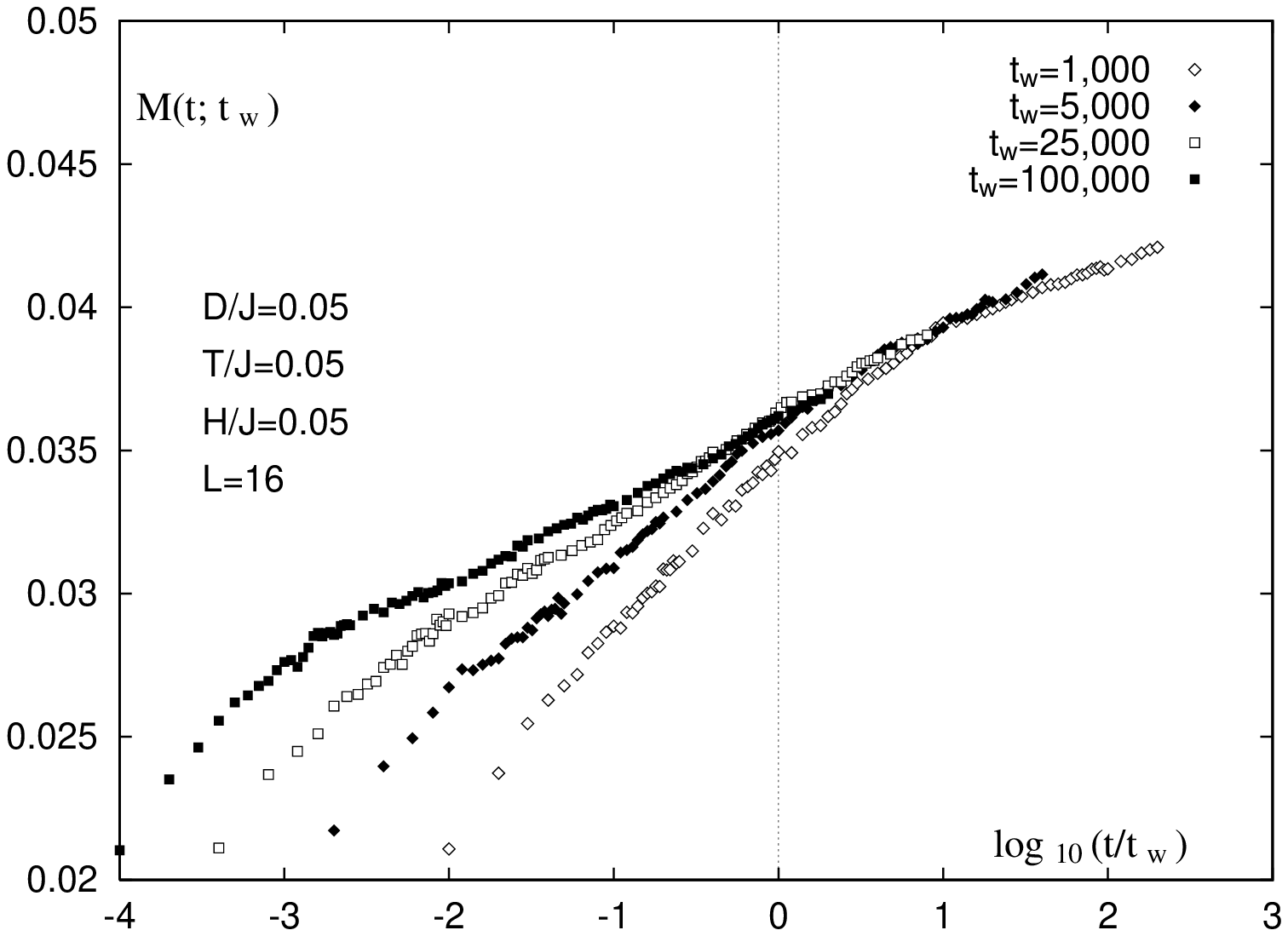}
}
\item{FIG.8~} Zero-field-cooled  magnetization
of an   anisotropic 3D Heisenberg spin glass with $D/J=0.05$
plotted versus $\log _{10}t$.
The field is  $h/J=0.05$ and
the temperature is $T/J=0.05$.
The lattice size is  $L=16$ 
averaged over 80 samples.
\smallskip

Experimentally,  thermoremanent magnetization (TRM) or 
zero-field-cooled (ZFC) magnetization is found to
show an approximate $t/t_w$-scaling
in the aging regime, with  small deviation from
the perfect scaling
in the direction of subaging [9]. Although this seems 
in apparent contrast to the present result,
it should be noticed that standard aging experiments 
have  been made by measuring the magnetic response,
not the autocorrelation. 
Recent numerical simulation by Yoshino {\it et al\/}  
revealed that, at least in the case of the 
SK model, 
TRM showed the subaging 
even when the spin correlation showed the superaging [20].
Thus, we also calculate the ZFC magnetization 
for an anisotropic model
with $D/J=0.05$:
After the initial quench, 
the system is  evolved in zero field during $t_w$ MCS. 
Then, an external field
of intensity $H/J=0.05$ is turned on and the subsequent growth 
of the magnetization
$M(t;t_w)$ is recorded. As can be seen from Fig.8, 
the data show the near $t/t_w$-scaling
in the aging regime $t>t_w$ where the 
spin-autocorrelation shows the superaging.
Thus, the observed tendency is roughly
consistent with experiments.  
It might be interesting to experimentally
investigate the aging properties of {\it spin correlations\/}
of Heisenberg-like magnets in search for  possible
superaging behavior.

\par
\bigskip\medskip
\noindent
\S 5. {\bf Summary}
\par\medskip
In summary, spin-glass and chiral-glass orderings in 
3D Heisenberg spin glasses are studied 
with and without random anisotropy 
by Monte Carlo simulations.     
The results are basically consistent with the chirality
mechanism: 
In the isotropic case,  clear evidence of the occurrence of a 
finite-temperature chiral-glass transition without the
conventional spin-glass order
is  presented both by equilibrium and off-equilibrium simulations.
Spin and chirality
show very different dynamical behaviors consistent with the
`spin-chirality separation'. While the spin 
autocorrelation exhibits only
an interrupted aging, the chirality autocorrelation
persists to exhibit a pronounced aging
effect reminiscent of the one observed
in the mean-field model. The universality
class of the chiral-glass transition appears to be 
different from that of the the standard
Ising spin glass.
In the anisotropic case, the off-equilibrium simulation indicates that
the spin shows the same asymptotic behavior as  
the chirality in the isotropic case, demonstrating
the `spin-chirality mixing' due to magnetic  anisotropy.

The authors are thankful to  I. Campbell, R. Orbach, H.Takayama, 
E.Vincent, A.P. Young, L.F.Cugliandolo, 
M.Ocio,  H.Rieger, K. Nemoto and H.Yoshino for useful
discussion. The numerical calculation was performed on the FACOM
VPP500 at the supercomputer center, ISSP, University of Tokyo and on
the HITACHI SR-2201 at the supercomputer center, University of Tokyo.

\bigskip\medskip
\noindent
{\bf References}
\medskip\par\noindent
\item{[1]} 
For  reviews on spin glasses, see, {\it e.g.,\/}
K. Binder and A.P. Young, Rev. Mod. Phys. {\bf 58},  801
(1986);
 K.H. Fischer and J.A. Hertz, {\it Spin Glasses\/}
Cambridge University Press (1991); 
J.A. Mydosh, {\it Spin Glasses\/} Taylor \& Francis (1993).

\item{[2]} N. Kawashima and A.P Young,
Phys. Rev. B{\bf 53}, R484 (1996). 

\item{[3]} K. Hukushima, H. Takayama and K.Nemoto, 
Int. J. Mod. Phys. {\bf 7}, 337 (1996). 

\item{[4]} J.A. Olive, A.P. Young and D. Sherrington,
Phys. Rev. B{\bf 34}, 6341(1986).

\item{[5]} F. Matsubara, T. Iyota and S. Inawashiro, 
Phys. Rev. Lett. {\bf 67},  1458  (l991).

\item{[6]} H. Kawamura, Phys. Rev. Lett. {\bf 68}  (l992) 3785;
Int. J. Mod. Physics {\bf 7},  345 (1996).

\item{[7]} H. Kawamura, J. Phys. Soc. Jpn. {\bf 64}, 26 (1995);
H. Kawamura and K. Hukushima, to appear in J. Mag. Mag. Mater.

\item{[8]} L. Lundgren, P. Svedlindh, P. Nordblad and O. Beckman,
Phys. Rev. Lett. {\bf 51}, 911 (1983).

\item{[9]} E. Vincent, J. Hammamm, M. Ocio, J-P. Bouchaud and L.F.
Cugliandolo, Sitges Conference on Glassy Systems, 
1996 

 (Springer, in press) cond-mat.9607224.

\item{[10]} J-P. Bouchaud, L.F. Cugliandolo, J. Kurchan and 
M. M\'ezard, in {\it Spin Glasses and Random Fields\/}, ed. by A.P.
Young, World Scientific (Singapore 1997) cond-mat.9702070.%

\item{[11]} L.F. Cugliandolo and J. Kurchan, Phys. Rev. Lett. {\bf 71},
173 (1993); J. Phys. A{\bf 27}, 5749 (1994); Phil. Mag. {\bf 71}, 
501 (1995).

\item{[12]} J.-O. Andersson, J. Mattsson and P. Svedlindh, Phys. Rev. 
B{\bf 46}, 8297 (1992); B{\bf 49}, 1120 (1994). 

\item{[13]} H. Rieger, J. Phys. A{\bf 26}, L615 (1993); H. Rieger,
B. Steckemetz and M. Schreckenberg, Europhys. Lett. {\bf 27}, 
485  (1994) ;
J. Kisker, L. Santen, M. Schreckenberg and H. Rieger, Phys. Rev.
B53, 6418 (1996).

\item{[14]} G. Parisi, F. Ricci-Tersenghi and J.J. Ruiz-Lorenzo,
J. Phys. A{\bf 29}, 7943 (1996). 

\item{[15]} A. Baldassarri,
cond-mat.9607162.

\item{[16]} H. Takayama, H. Yoshino and K. Hukushima, 
J. Phys. A{\bf 30}, 3891 (1997).

\item{[17]} E. Marinari, G. Parisi, D. Rossetti,
cond-mat.9708025.

\item{[18]} K. Hukushima and K. Nemoto, J. Phys. Soc. Jpn.
{\bf 65}, 1604 (1996).

\item{[19]} J. Vilain, J. Phys. C{\bf 10}, 4793 (1977);
C{\bf 11}, 745 (1978).

\item{[20]} H. Yoshino, K. Hukushima and H. Takayama, 
Prog. Theor. Phys. Suppl. {\bf 126}, 107 (1997).

%
%
%

\end